\documentclass[aps, prl, reprint]{revtex4-1}

\usepackage{amsmath}
\usepackage{graphicx}
\usepackage{float}
\usepackage{CJK}
\usepackage{eqnarray}

\begin{document}

\title{Supercurrents in chiral channels originate from upstream information transfer: a theoretical prediction}
\date{\today}
\author{Xiao-Li Huang  }
\affiliation{Kavli Institute of NanoScience, Delft University of Technology}
\author{Yuli V. Nazarov}
\affiliation{Kavli Institute of NanoScience, Delft University of Technology}

\begin{abstract}
It has been thought that the long chiral edge channels cannot support any supercurrent between the superconducting electrodes. We show theoretically that the supercurrent can be mediated by a non-local interaction that facilitates a long-distance information transfer in the direction opposite to electron flow. We compute the supercurrent for several interaction models, including that of an external circuit.  
\end{abstract}

\maketitle

Proximity effect in normal metal-superconducting structures is known for a long time \cite{DdG, McMillan} but still is a subject of intense theoretical and experimental research \cite{CCBCMDSR, MeGrenoble}. The most prominent manifestation of proximity effect is a supercurrent flowing through a normal metal between distant superconducting electrodes. The interesting feature of the effect is that the induced superconducting correlations persist in a normal metal on  
long distances. The distance even diverges at energies close to Fermi level, $\epsilon \equiv E-E_F \to 0$, $L \simeq v_F/|\epsilon|$ for ballistic structures with typical electron velocity $v_F$.

In Quantum Hall regime, the conducting electrons are restricted to the quantized transport channels at the structure edge\cite{Halperin}. Importantly, these channels are chiral: the electrons propagate in one direction only. Superconducting leads connected to the edge modes may induce the proximity effect. Interesting Andreev reflection phenomena \cite{Takagaki, Chtchelkatchev} and the supercurrent in chiral channels \cite{MZ, SL, vOAB}have been thoroughly investigated. Notably, it was shown that the supercurrent carried by a chiral channel requires the closing of the channel and is inversely proportional to the full perimeter of the QH sample. Therefore there seem to be no current in the situation when this perimeter is macroscopically long, for instance, in the situation given in Fig. 1a. An heuristic explanation is that the supercurrent is due to the bouncing of electrons and Andreev-reflected holes between the superconducting electrodes. In a chiral channel, both electrons and holes move in the same direction and no bouncing can occur unless a particle encircles the perimeter of the whole macroscopic sample. If there were transport channels propagating in the opposite direction, we would have a current of the scale $ev_F/L$, $L$ being the distance between the superconducting electrodes. The absence of the supercurent seems a simple but fundamental property of the chiral channels. It is not affected by local electron-electron interactions in the channel that can be easily taken into account in a framework of a Luttinger-like model \cite{Lutt}.

In this Letter we show that the supercurrent in a chiral channel can be induced by a non-local interaction that potentially provides an information flow in upstream direction, that is, opposite to the propagation direction of the electrons. 
  
We compute the supercurrent for several interaction models and demonstrate that the current is limited by a typical information transfer rate. The effect persists in ground state where no actual event of information transfer takes place: rather, the supercurrent indicates {\it potential} for such events. A transport mechanism based on information transfer is rather exotic for electrons and its experimental observation would be rewarding. We consider a situation of special experimental relevantce where the interaction is arranged by means of an external electric circuit.

\begin{figure}[h]

\includegraphics[width=.70\linewidth]{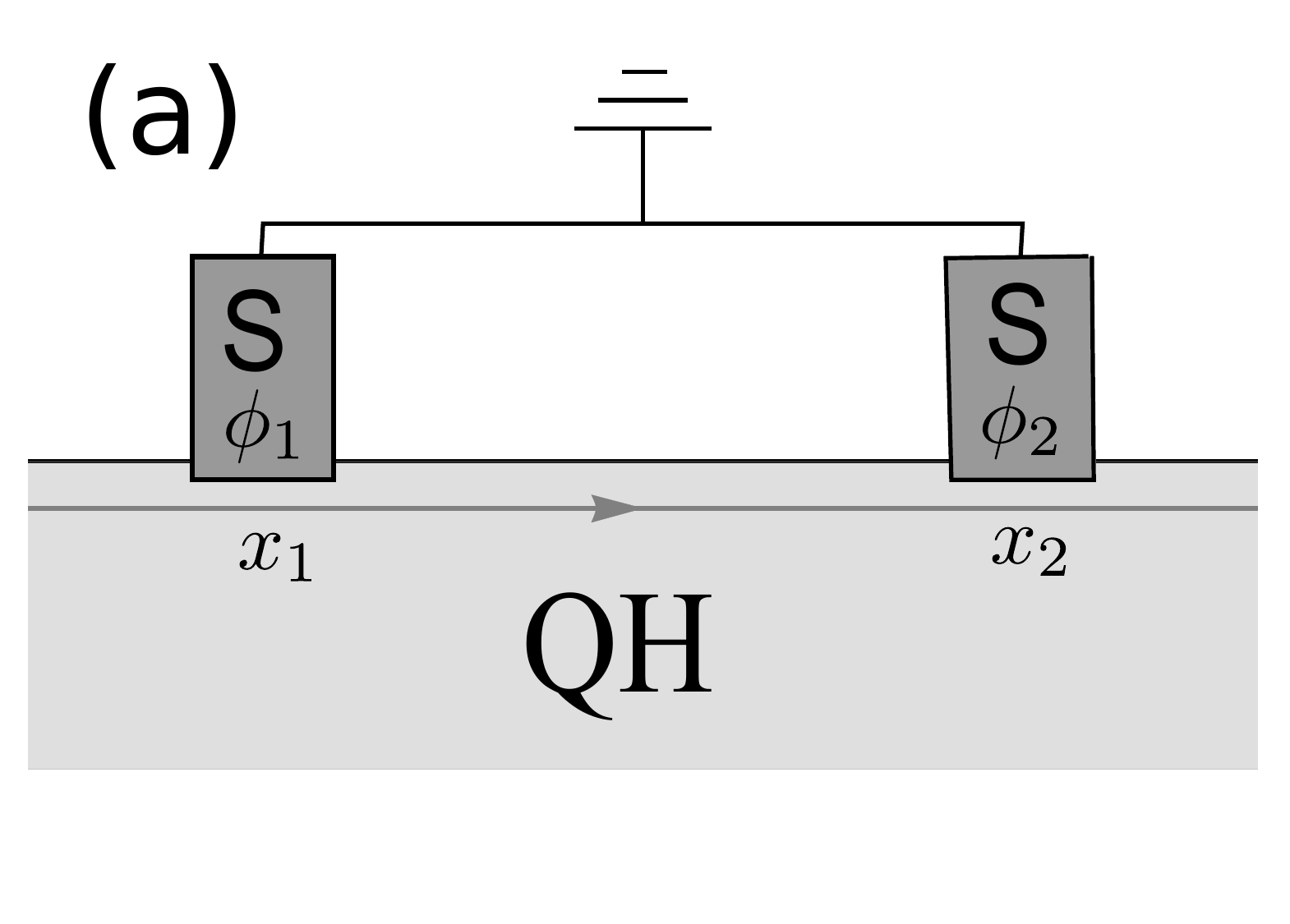}	\\
\includegraphics[width=.45\linewidth]{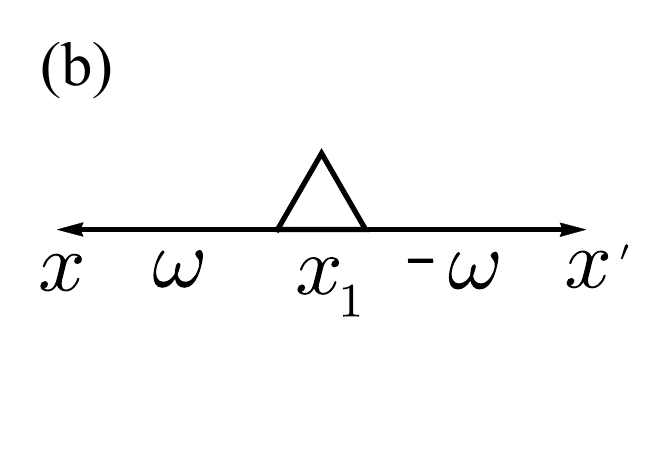}	\qquad
\includegraphics[width=.45\linewidth]{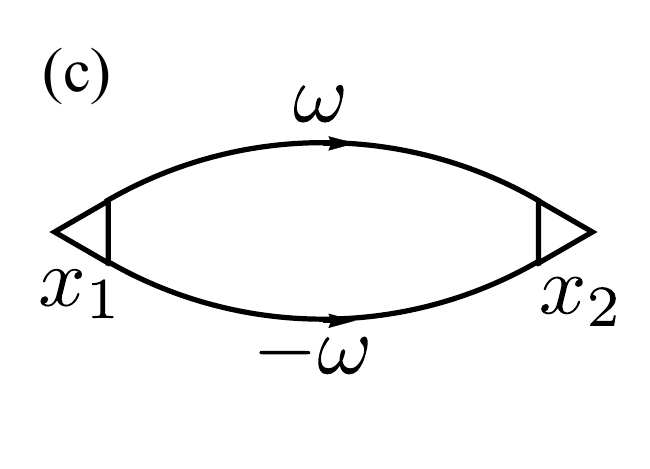}	

\caption{(a) The basic setup: two superconductor leads biased at phase difference $\phi =\phi_2 - \phi_1$ are attached to the edge of a quantum Hall sample; (b) The lowest order anomalous Green's function; (c) The non-interacting lowest order contribution shown vanishes for chiral channels.}
\label{fig: figure1}
\end{figure}

 The primary setup under consideration is shown in Fig.(\ref{fig: figure1}a). Assuming the macroscopically large QH sample, we consider an infinite 1D chiral channel at the sample edge. For simplicity, we concentrate on a single spin-degenerate channel: more channels do not change the results qualitatively. Two superconducting electrodes separated by distance $L = x_2- x_1$ are in contact with the channel, the contact length being $\ll L$. They are kept at the superconducting phase difference $\phi=\phi_2-\phi_1$. 
We assume low energies at the scale of Landau level separation. In this limit, the electron states in the channel can be described by a simple linearized Hamiltonian:
\begin{equation}
	H_0	= - i v_F 	\int \text{d}x \, \psi^\dagger(x) \partial_x \psi(x).
\end{equation}
The normal-electron Green's function in Matsubara representation  is  explicitly chiral,
\begin{equation}
G(i\omega_m,x-x')	=  -\frac{i {\rm sgn}(\omega)}{v_F} e^{-\frac{\omega}{v_F}(x-x')} \theta(\omega(x-x'))
\label{eqn:GreenF1}
\end{equation}
it extends to the right(left) for positive (negative) $\omega$ and is zero on the left(right). The lowest-order anomalous Green function $F_\omega(x,x')$ induced by a superconducting pairing at point $x_1$ (Fig. 1b)  encompasses the normal Green's functions at opposite frequencies
\begin{equation}
F_\omega(x, x')= \Delta(x_1) G(i \omega, x,x_1) G(-i\omega, x',x_1)
\end{equation}
This describes the superconducting correlations that are essentially non-local and, owing to chirality, vanish at the same point $x=x'$.  The superconducting current is expressed through the phase-dependent energy correction. This one could emerge from the transfer of the superconducting correlations to the second contact, $\simeq \sum_\omega F(x_2,x_2) \Delta(x_2)$ (Fig. 1c) yet it vanishes since the correlations vanish at the same point. The main point of this Letter is that a non-local interaction can change this. Let us consider a diagram shown in Fig. 2. Here, at positive $\omega$ the correlations propagate from $x_1$ to the separated points $x >x_1$ and $x_1>x'$. The non-local interaction between these distant points can flip the frequency sign of the electron line, $\omega'<0$,
so the correlations move in opposite direction to meet at the point $x_2$. This works provided $x'<x_1<x_2<x$. We see that the interaction should connect the region to the left from both electrodes with the region to the right of both.

\begin{figure}[h]

\includegraphics[width=.70\linewidth]{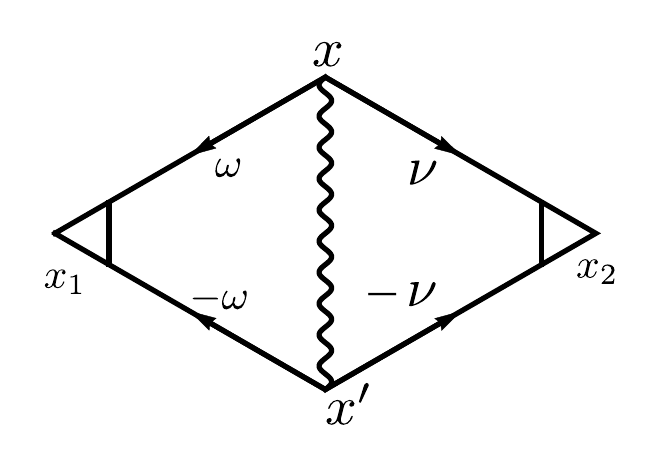}

\caption{Non-vanishing contribution to superconducting current. The wavy line represents interaction. By virture of chirality, the points $x,x'$ are on opposite sides of both $x_1$ and $x_2$. This is only possible if $x'<x_1<x_2<x$ and $\omega$ and $\nu$ are of opposite signs.}
\label{fig:figure2}
\end{figure}
   
%
%
%

Let us proceed with the evaluation of the current. We start with non-interacting Green's functions in the system, those are easy to evaluate in all orders in pairing potential that is incorporated in a form of a unitary transformation in Nambu space. This transpormation relates the elecron-hole amplitudes before and after the electrode and reads $\hat{U} = \exp(-i \int dx \hat{\Delta}(x)/{v_F})$, $\hat{\Delta}\equiv(0, \Delta ; \Delta^*, 0)$ where integration is taken in the vicinity of an electrode. Following \cite{Beenakker}, we can conveniently parametrize this matrix with the probability $p$ of Andreev electron-hole conversion at the corresponding contact, $\hat{U}_{1,2} =(\sqrt{1-p_{1,2}}, -i \sqrt{p_{1,2}} e^{i\phi_{1,2}},-i \sqrt{p_{1,2}} e^{-i\phi_{1,2}}, \sqrt{1-p_{1,2}})$
When superconductor leads are brought into proximity of the edge of the QH sample as in Fig.(\ref{fig: figure1}a), the induced pairing potential results in an unitary transformation on the . In the usual Nambu spinor formulation, for $x$ and $x'$ lying on different sides of the superconducting contacts, 
\begin{eqnarray}
	G(i\omega,x,x')	&=&	- \frac{i}{v_F}		e^{-\frac{\omega}{v_F}	(x-x')}	[	\theta(\omega) \theta(x-x') 		\hat{U}_2 \hat{U}_1	\nonumber \\	&& -	\theta(-\omega) \theta(x'-x)	\hat{U}_1^{\dagger} \hat{U}_2^{\dagger}].
\label{eqn:GF2}
\end{eqnarray}

 We evaluate correction to the energy brought by interaction and differentiate it with respect to $\phi$ to obtain the current,
 \begin{eqnarray}
 I(\phi) &=& -8  e R \sqrt{p_1 p_2 (1- p_1)(1- p_2)}\sin\phi;\\
 R &\equiv& (\frac{T}{v_F})^2\sum_{\omega,\omega'} \Theta(-\omega\omega') \int_{-\infty}^{x_1} dx \int_{x_2}^{\infty} dx' \nonumber \\&V&(\omega-\omega'; x,x') e^{|\omega-\omega'|(x-x')/v_F}. \label{eq:relation}
 \end{eqnarray}
 Here we have not yet specified the form of interaction $V(\nu; x,x')$. We see that the current assumes a usual sinusoidal Josephson phase dependence and is proportional to $\sqrt{p_1 p_2 (1- p_1)(1- p_2)}$ which indicates the current comes about the interference of two processes: i. an electron propagation with Andreev reflection in superconductor 1 and no Andreev reflection in superconductor 2,  and  the propagation with no reflection in superconductor 2 and Andreev reflection in superconductor 2. All details of the junctions are incorporated in $p_{1,2}$ while the coefficient $R$ characterizes the interaction in the setup.  Further we evaluate the coefficient $R$ for various interaction setups and prove its relation to the rate of the upstream information transfer. 

We start with a rather artificial but instructive setup. Let us consider a harmonic oscillator with eigenfrequency $\omega_b$ that is coupled to the edge states in two points $x_{3,4}$ The coupling is described with ($\hat{n}(x)\equiv\sum_{\sigma} psi^\dagger_\sigma	(x)\ \psi_\sigma(x)$)
\begin{equation}
	H_I	=	(\alpha_3 \hat{n}(x_3) + \alpha_4 \hat{n}(x_4))	(\hat{b} + \hat{b}^\dagger)	.
\end{equation}
Owing to the coupling, the quanta of the oscillator can be absorbed by the edge states with the rates $\Gamma_{3,4} =\alpha_{3,4}^2 \varepsilon_b/v_F^2$. The oscillator provides a channel of upstream information exchange whereby an excitation at the point $x_4$ is absorbed and transferred to the upstream point $x_3$. The information flow rate through the oscillator is limited by the emission/absorption rates and can be estimated as ${\rm min}(\Gamma_3,\Gamma_4)$. 

Let us look at the superconducting properties assuming $x_3<x_1<x_2<x_4$. The oscillator provides an effective electron-electron interaction ($x>x'$):
\begin{equation}
	V(x,x',\nu)=\frac{\alpha_3 \alpha_4  \omega_b}{\omega_b^2 + \nu^2} \delta (x-x_3) \delta(x'-x_4)
\label{eqn:e-ph-interaction}
\end{equation}
Making use of the relation (\ref{eq:relation}), and integrating over the frequencies, we arrive at the coefficient $R$ charaterizing the current,
\begin{equation}
R = \frac{1}{2 \pi} \sqrt{\Gamma_3 \Gamma_4} C;
\end{equation}
 where the dimensionless coefficient $C$ in two opposite limits $\omega_b \ll v_F/L$ and $\omega_b \ll v_F/L$ is evaluated as $C = \ln (\omega_b L/v_F)$ and $C=(v_F/L\omega_b)^2$ respectively. 
 We see that for $\omega_b \ll v_F/L$ the coefficient $R$ is of the order of the information transfer rate. This correspondence is not exact, as seen from different dependence on $\Gamma_3/\Gamma_4$ ratio. This is not surprising since, in distinction from information transfer, no real events are associated with the supercurrent that results from quantum interference. However, such correspondence is remarkable even on a qualitative level. As seen from Eq. \ref{eq:relation}, the relevant frequency window for supercurrent formation is limited by $v_F/L$. This explains suppression of $R$ at $\omega_b \gg v_F/L$: the oscillator cannot efficiently transmit such low frequencies.

A Quantum Hall sample is always mounted on a substrate. This makes electron-phonon interaction a default mechanism for a long-range upstream information flow: an electron-hole pair can be converted to a phonon that propagates upstream and is absorbed there. We describe the electron-phonon interaction by the Hamiltonian
\begin{equation}
	H_{e-ph}=A \sum_{\vec{q}} 	\frac{i \vec{q}}{\sqrt{2 \rho \,  \omega_{\vec{q}} V}}
			 \int \text{d} \vec{x} \, 	e^{i \vec{q} \cdot \vec{x}}		\hat{n}(\vec{x})
			 (		a_{\vec{q}} 	+	a^\dagger_{-\vec{q}}	)	.
\label{eqn: e-ph Hamiltonian}
\end{equation}
Here $\rho$ is the substrate density, ${\cal V}$ in the normalization volume, $\vec{q}$ the phonon wave vector, and $\omega_{\vec{q}}=c |\vec{q}|$ with $c$ being the sound velocity. For electrons in the edge channel, $\vec{x}$ is one-dimensional. This results in the following long-range electron-electron interaction 
\begin{equation}
	V(x, x', \nu)	= -	\frac{A^2}{2 \rho {\cal V} }
					\sum_{\vec{q}}  		
					 \frac{\vec{q}^2}{\nu^2 + \omega^2_{\vec{q}}}	
					 e^{i \vec{q} \cdot (\vec{x}-\vec{x}')}
					\label{eqn: effective interaction}
\end{equation}
Let us note the analogy with the previous setup: each phonon mode is in fact an oscillator that is coupled to the electrons both upstream and downstream of the superconducting contacts. 

The strength of the interaction is convenient to express in terms of the electron relaxation rate, that is proportional to $\epsilon^3$, $\epsilon$ being the electron energy above the Fermi level,
\begin{equation}
	\Gamma(\epsilon)	=	\frac{A^2}{12 \pi \rho \, v_F \, c^4}	(\epsilon) ^3	\equiv\frac{\text{d} \Gamma}{\text{d} \epsilon^3 } 	\epsilon^3 .
	\label{eqn: relaxtion rate}
\end{equation}

Integrating over all oscillators, we arrive at superconducting current defined by 
\begin{equation}
	R	= \frac{3}{16 \pi^2}		\frac{\text{d} \Gamma}{\text{d} \epsilon^3}		\frac{c}{v_F}	\left(\frac{c}{L}\right)^3 .
	\label{eqn: energy final 2}			
\end{equation}
The typical energies involved in the integration are of the order of inverse sound propagation time between the superconducting junctions, $c/L$. To estimate the typical information transfer rate, let us consider electrons excited to these energies. The phonon information transfer rate is the number of relevant excitations times the relaxation rate of a single excitation, 
$(\text{d} \Gamma/\text{d} \epsilon^3) (c/L)^3$. The relevant excitations are at the space scale $\simeq L$, so their number is $(c/v_F)\ll 1$. This reproduces $R$ by order of value.

However, for realistic structures, the intrinsic electron-phonon effect is fairy small albeit intrinsic.
For typical GaAs parameters, 
$c/L = 10^{10} \text{Hz}$, $c/v_F=10^2$ \cite{Altimiras} and  
we estimate 
$d\Gamma/d\epsilon^3 \simeq \omega^{-2}_D\simeq 5\cdot 10^{-28} \text{Hz}^{-2}$. 
All this gives $R \simeq 0.1 \text{Hz}$, and the corresponding current is truly unmeasurable. Аt low energies the electron-electron interaction, that is, interaction with electricity fluctuations, is more important for relaxation than the phonons \cite{Altimiras}. However, it is not obvious that electon-electron interaction alone can provide the upstream information transfer required. For instance, the edge magnetoplasmons transfer information only downstream.

There is a simple way to circumvent this: one can embed the QHE edge into an external electric circuit that will transfer the electric signals upstream. This is the last setup that we consider. It has advantages of tunability since the strength of the long-range interaction is determined by the circuit parameters. As we will see, it also provides large values of the supercurrent.

\begin{figure}[h]

\includegraphics[width=.5\linewidth]{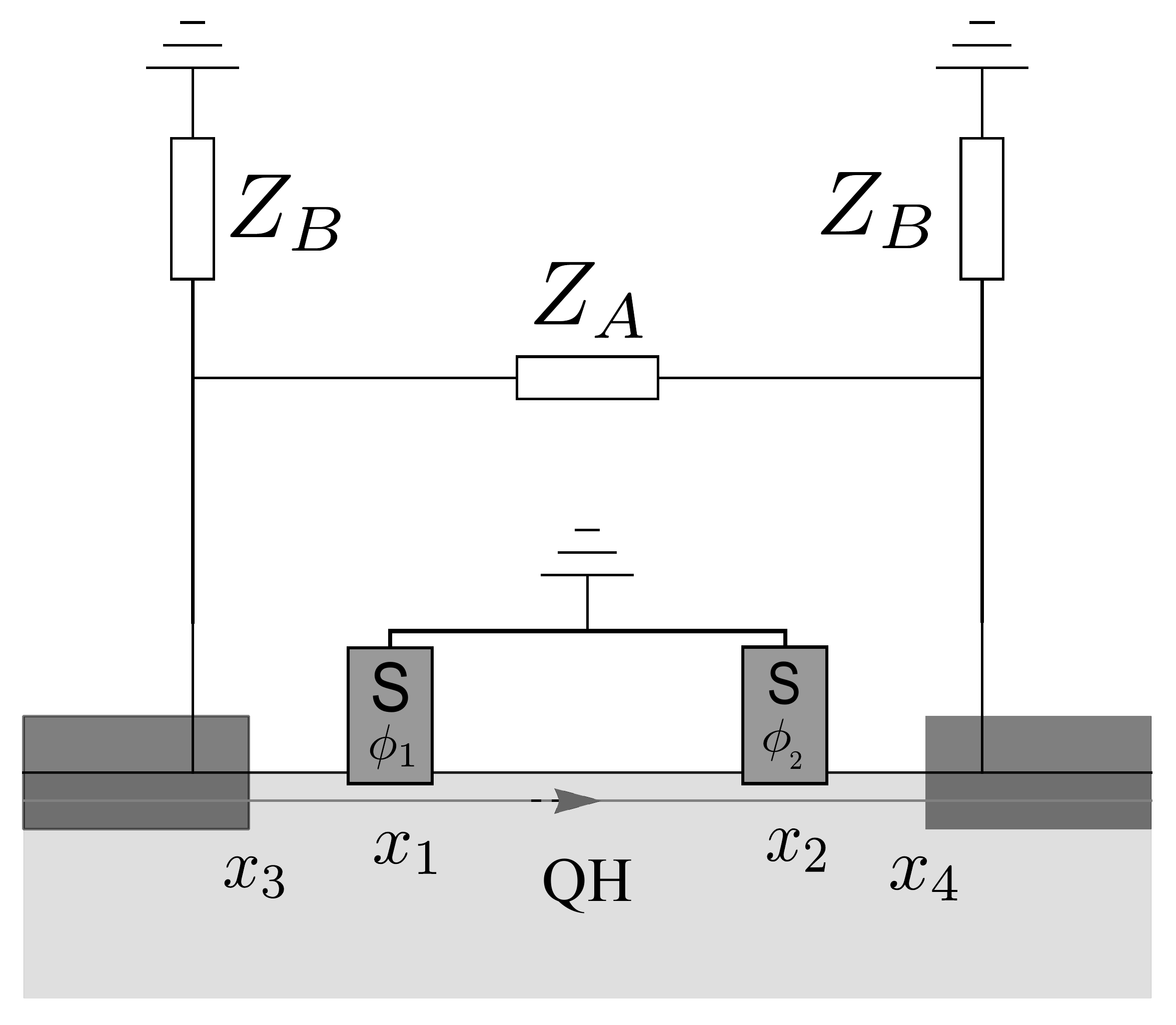}

\caption{The QH edge with superconducting contacts included into an external circuit characterized by (frequency-dependent) impedances $Z_{A,B}$. The superconducting current between the terminals 1,2 is proportional to the cross-impedance $Z_{34}$.
Dark-grey: metal contacts covering QHE structure.}
\label{figExternal}
\end{figure}

To describe the connection of the edge with an external circuit, we cover it with two metallic electrodes that are spread at $x<x_3$ and $x>x_4$ respectively, ($x_4-x_3=\tilde{L}$) and are characterized by fluctuating voltages $\hat{V}_{3,4}$. It is convenient to make a guage transform introducing $\varphi_{a,b}(t) = e \int^t_{-\infty} \hat{V}_{ab}$ that is a phase shift induced by a corresponding voltage. With this, the interaction with the external circuit is local,
\begin{equation}
H_\varphi = - v_F(\hat{\varphi}_3 \hat{n}(x_3) - \hat{\varphi}_4 \hat{n}(x_4)
\end{equation}
and is similar to that in the setups considered. The effective interaction is expressed in terms of the correlator of the phases,
\begin{equation}
V(x, x', \nu) = \frac{v_F^2}{2}  \langle \varphi_{3}(\nu) \, \, \varphi_{4}(-\nu) \rangle \delta (x-x_a) \delta(x'-x_b)
\end{equation}
that is related to the frequency-dependent {\it cross-impedance} $Z_{34}(\nu)$ between the leads 3 and 4,
\begin{equation}
\langle \varphi_{3}(\nu) \, \, \varphi_{4}(-\nu) \rangle = \frac{Z_{34}(\nu)}{\nu}
\end{equation}
For the circuit in Fig. \ref{figExternal}, $Z_{34}=Z_B^2/(Z_A+2Z_B)$. We obtain the current coefficient from Eq. \ref{eq:relation}, 
\begin{equation} 
\label{eq:withimpedance}
R	=	\frac{e^2}{\pi^2}  \int_0^{\infty} \text{d} \omega  \,	e^{-\frac{\omega}{v_F} \tilde{L}} Z_{34}(\omega)
\end{equation}

A simple relation is obtained for a frequency-independent cross-impedance:
\begin{equation}
R = \frac{e^2}{\pi^2} \frac{v_F}{\tilde{L}} Z_{34}
\end{equation} 

This also can be interpreted as a potential information transfer rate, given the bandwidth $v_F/\tilde{L}$ and the fraction of information transferred upstream defined as the ratio of impedances $Z_{34}/R_Q$, $R_Q \equiv \pi\hbar/e^2$ being the self-impedance of the QHE edge. Upon increasing the impedance of the external circuit to the values of the order $R_Q$, this fraction becomes of the order of $1$ and the effect is maximized up to $R$ of the order of the bandwidth.

In conclusion, we have demonstrated theoretically that supercurrent can exist in long chiral channels. In distinction from all known mechanisms of supercurrent, it essentially requires interaction. Moreover, it requires a special kind of non-local interaction that connects points that are downstream of the superconducting electrodes to the points upstream of those. This connection is no galvanic: it is not the charge that is transferred upstream but rather the information about the charge transfer. We agrue that the maximum value of the supercurrent is accocsiated with the rate of upstream information transfer, at least at qualitative level. Even at this level, this relation is rather intriguing since the supercurrent is a property of the ground state where no process associated with information transfer can occur. This suggests that the supercurrent can probe the potential for information transfer without actually transferring the information. This may be useful in the context of defining quantum information flows \cite{Iflow1,Iflow2}. On practical side, this property of the supercurrent makes it feasible to check if in more complex QHE states all edge channels actually flow in the same direction. \cite{Kane} It is feasible to observe the effect experimentally. The traditional difficulties of good contact between metals and 2D gas can be circumvented if utilitizing QHE edge channels in graphene \cite{graphene, graphene2}. The best setup is likely one with the external circuit provided the impedances involved can be controlled on-chip proving the scaling predicted by Eq. \ref{eq:withimpedance}. Here we present the results at vanishing temperature. We expect the temperature to start playing a role at $k_B T \simeq v_F/L$ at the current to decrease exponentially, $ R \propto \exp(-k_B T L/v_F)$, at $k_B T \simeq v_F/L$.

This work is part of the research programme of the Foundation for Fundamental Research on Matter (FOM), which is part of the Netherlands Organisation for Scientific Research (NWO). The authors acknowledge useful discussions with A. Akhmerov and A. Yacoby.

\bibliography{Draft}

\end{document}